\documentclass[lettersize,journal]{IEEEtran}
\usepackage{amsmath,amsfonts}
\usepackage{algorithmic}
\usepackage{algorithm}
\usepackage{array}
\usepackage[caption=false,font=normalsize,labelfont=sf,textfont=sf]{subfig}
\usepackage{textcomp}
\usepackage{stfloats}
\usepackage{url}
\usepackage{verbatim}
\usepackage{graphicx}
\usepackage{cite}
\usepackage{aliascnt}
\usepackage{amsthm}
\usepackage{supertabular}
\usepackage{placeins}
\usepackage{afterpage}
\usepackage[labelsep=period]{caption}
\usepackage{ragged2e}

\usepackage{aliascnt}
\usepackage{caption}

\newtheorem{thm}{Theorem}

\newtheorem{lem}{Lemma}

\newtheorem{prop}{Proposition}

\newcommand{\norm}[1]{\|#1\|}

\usepackage{ltablex}
\keepXColumns

\begin{document}

\title{Residual-Evasive Attacks on ADMM in Distributed Optimization}

\author{Sabrina Bruckmeier \thanks{Sabrina Bruckmeier is with the Graduate School of Mathematics, ETH Zurich, Zurich, Switzerland}, Huadong Mo \thanks{Huadong Mo is with the School of Systems and Computing, University of New South Wales, New South Wales, Australia}, ~\IEEEmembership{Senior Member,~IEEE,} and James Ciyu Qin \thanks{James Qin is with the Reliability and Risk Engineering Laboratory, Institute of Energy and Process Engineering, Department of Mechanical and Process Engineering,  ETH Zurich, Zurich, Switzerland} ~\IEEEmembership{Member, ~IEEE}}
        % <-this % stops a space
% <-this % stops a space

%\thanks{Manuscript received x; revised x.}}


%
%\markboth{IEEE Transactions on Smart grid,~Vol.~xxx, No.~x, month~x}%
%{Shell \MakeLowercase{\textit{et al.}}: Residual-Evasive Attacks on ADMM in Distributed Optimization}


\maketitle

\begin{abstract}
This paper presents two attack strategies designed to evade detection in ADMM-based systems by preventing significant changes to the residual during the attacked iteration. While many detection algorithms focus on identifying false data injection through residual changes, we show that our attacks remain undetected by keeping the residual largely unchanged. The first strategy uses a random starting point combined with Gram-Schmidt orthogonalization to ensure stealth, with potential for refinement by enhancing the orthogonal component to increase system disruption. The second strategy builds on the first, targeting financial gains by manipulating reactive power and pushing the system to its upper voltage limit, exploiting operational constraints. The effectiveness of the proposed attack-resilient mechanism is demonstrated through case studies on the IEEE 14-bus system. A comparison of the two strategies, along with commonly used naive attacks, reveals trade-offs between simplicity, detectability, and effectiveness, providing insights into ADMM system vulnerabilities. These findings underscore the need for more robust monitoring algorithms to protect against advanced attack strategies.
\end{abstract}

\begin{IEEEkeywords}
ADMM, Cybersecurity, Optimal Power Flow, Distributed Optimization, Data Manipulation.
\end{IEEEkeywords}

\section{Introduction}
\IEEEPARstart{T}{he} Alternating Direction Method of Multipliers (ADMM) has become a widely used optimization algorithm across various fields, including power systems, due to its scalability and effectiveness in solving large-scale, distributed optimization problems. As the adoption of ADMM continues to grow, its vulnerability to data manipulation attacks has become a significant concern. These attacks can undermine the integrity of the optimization process, potentially leading to compromised system performance and security. Consequently, there has been a growing focus on developing robust detection algorithms to identify and mitigate such threats. This section provides an overview of the most relevant detection techniques in the literature.

Alkhraijah et al. \cite{DetectingSharedDataManipulationInDistributedOptimizationAlgorithms} introduced two detection mechanisms, Convergence Consistency (CC) and Solutions Consistency (SC), which identify data manipulation by monitoring convergence trajectories and consistency across iterations. In \cite{Alkhraijah2022AnalyzingMD}, this analysis was extended to the Auxiliary Problem Principle (APP) algorithm, proposing a neural network-based framework trained on shared variable mismatches, highlighting the potential of data-driven approaches in distributed optimization.

Residual-based detection methods play a pivotal role for identifying False Data Injection Attacks (FDIAs). Obata et al. \cite{obata2023detection} used intermediate residuals from the ADMM process to detect attacks early by identifying sudden spikes, while Liao and Chakrabortty \cite{RoundRobin} introduced the Round-Robin ADMM (RR-ADMM) algorithm, tracking spiked values to identify malicious agents. Both approaches rely on detecting abnormalities to flag malicious behavior.
These mechanisms address basic attack strategies, such as naive false data injection, constant offset attacks, and random noise injection, forming a foundation for understanding ADMM’s vulnerabilities. Building on this, our work investigates how residual-based detection methods like CC and SC can be bypassed through a novel attack strategy.

Zhai et al. \cite{Zhai} proposed a robust optimization framework to address wind power uncertainty in integrated power and gas systems, utilizing Linear Decision Rules (LDRs) and Automatic Generation Control (AGC) to balance robustness and scalability. Similarly, Duan et al. \cite{Duan} introduced a resilient DC Optimal Power Flow (DC-OPF) algorithm that combines Bayesian reputation functions with information estimation to detect and mitigate data integrity attacks. These methods highlight the need for dynamic, adaptable frameworks to handle sophisticated threats. Xu et al. \cite{ADMMBasedOPF} advanced the discussion on cyberattack resilience, demonstrating the effectiveness of Artificial Neural Network-based mechanisms to mitigate time-delay and data manipulation attacks, showcasing the potential of combining optimization techniques with machine learning to enhance security, particularly in scenarios requiring rapid response. Li et al. \cite{FDIAttacksInSmartGrids} employed federated deep learning with Transformer models to achieve high detection accuracy while maintaining data privacy.

In the broader context of machine learning applications, Xie et al. \cite{ReviewOfMLinPowerSystem} reviewed advanced machine learning methods, including deep reinforcement learning (DRL), convolutional neural networks (CNNs), and ensemble learning, for tasks like stability assessment and outage prediction. They highlighted DRL’s ability to integrate perception and decision-making, while CNNs excelled in analyzing high-dimensional data for stability issues. Similarly, Tuyizere and Ihabwikuzo \cite{tuyizere2023machine} showed that Random Forest models effectively detect and classify power system disturbances, distinguishing between natural events and cyberattacks with high accuracy. Chatterjee et al. \cite{ReviewOfCyberAttacks} conducted a comprehensive review of cyberattacks on power systems, focusing on their mechanisms, impacts, and vulnerabilities across state estimation, AGC, energy markets, and particularly interesting for this paper voltage control. They highlighted attack strategies such as FDIA on state estimation and data integrity attacks on Load Tap Changing (LTC) transformers, Flexible AC Transmission System (FACTS) devices, and Locational Marginal Prices (LMPs). The review emphasized the challenges of detecting coordinated, well-crafted attacks and the urgent need for robust cybersecurity measures and real-time detection systems. 

\begin{figure}[!]
\centering
  \includegraphics[width=\linewidth]{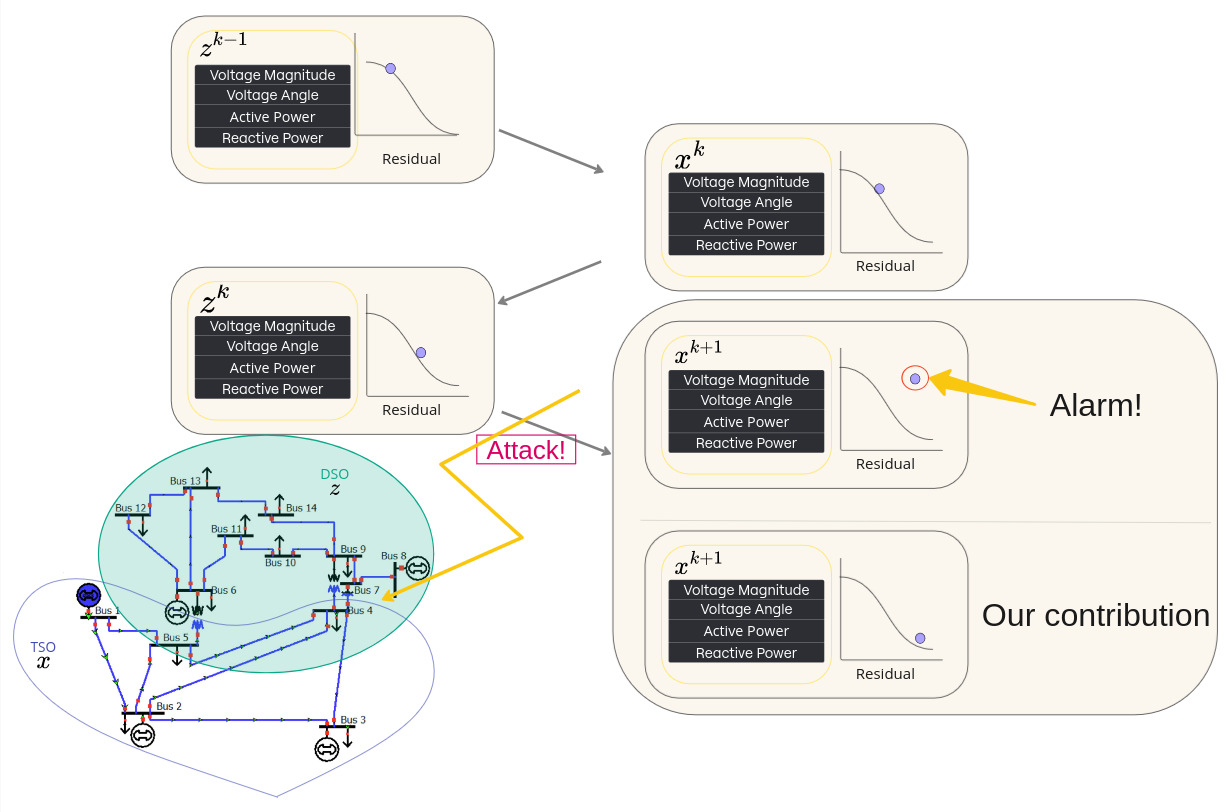}
  \caption{Our contribution: Evasion through minimal residual changes.}
  \label{fig: Process}
\end{figure}

Unlike conventional attacks that cause detectable perturbations, our approach manipulates the system without altering the primary residual in the attacked iteration---a behavior we refer to as residual-evasive ---thereby evading standard monitoring strategies. The main contributions of this paper as illustrated in Fig. \ref{fig: Process} are three residual-evasive attacks, which underscore the vulnerability of distributed OPF algorithms.
\begin{enumerate}
\item We demonstrate that even random attack vectors can bypass residual-based detection.
\item We enhance the effect of random attacks by aligning them with directions orthogonal to the system's projected trajectory, using Gram-Schmidt orthogonalization to maximize their deviation while preserving stealth.
\item We develop targeted residual-evasive attacks that steer the system toward specific undesirable states, with a focus on voltage control---a critical and financially incentivized function in power system operations—--by exploiting vulnerabilities in the ADMM where natural fluctuations mask their effects. 
\end{enumerate}

The remainder of this paper is organized as follows. Section \ref{sec: Background} provides background on ADMM and explains how it applies to OPF. Section \ref{sec: Vulnerabilities} explores vulnerabilities in ADMM and introduces the notion of residual-evasive attacks, corresponding to our first contribution. Section \ref{sec: Goal - Oriented Attacks in ADMM} presents the third and second contributions in order: we first develop targeted attacks that steer the system toward specific states, followed by optimized random attacks that maximize deviation while remaining undetected. Section \ref{sec: Simulation Results} evaluates the effectiveness of the proposed strategies through a comprehensive case study on the IEEE 14-bus system. Finally, Section \ref{sec: Conclusion} concludes the paper and discusses future research directions, including ways to strengthen ADMM against such vulnerabilities.

\section{Background}
\label{sec: Background}
ADMM, introduced in the 1970s by Glowinski and Marrocco \cite{ADMMGlowinski} and Gabay and Mercier \cite{ADMMGabay}, is an optimization algorithm designed to solve large-scale convex problems. It combines the decomposability of dual ascent methods with the convergence robustness of the method of multipliers, making it suitable for distributed optimization in fields such as machine learning, signal processing, and power systems. The optimization problem it solves is formally written as 
\begin{equation}
\label{ADMM Problem}
\min f(x)+g(z) 
\text{ s.t. } Ax + Bz = c
\end{equation}
where $f(x)$ and $g(z)$ are convex functions, and $x$ and $z$ are optimization variables subject to linear constraints defined by $A$, $B$ and $c$. ADMM solves this using an augmented Lagrangian, introducing a penalty alongside the usual Lagrangian for constraint violations:
\begin{align}
\label{ADMM Lagrangian}
&L_{\rho}(x,z,\lambda) \\
=& f(x) + g(z) 
+ \lambda^T(Ax+Bz-c) \notag 
+ \frac{\rho}{2} \norm{Ax + Bz - c}_2^2
\end{align}
where $\lambda$ is the dual variable (Lagrange multiplier) and $\rho >  0$ is a penalty parameter. The algorithm proceeds iteratively with updates for $x$, $z$, and $\lambda$: \\ 
\textbf{Update \(x\):}
\begin{equation*}
x^{k+1} = \arg\min_x \left( f(x) + \frac{\rho}{2} \|Ax + Bz^k - c + \lambda^k\|_2^2 \right),
\end{equation*}
\textbf{Update \(z\):}
\begin{equation*}
z^{k+1} = \arg\min_z \left( g(z) + \frac{\rho}{2} \|Ax^{k+1} + Bz - c + \lambda^k\|_2^2 \right),
\end{equation*}
\textbf{Update \(\lambda\):}
\begin{equation*}
\lambda^{k+1} = \lambda^k + Ax^{k+1} + Bz^{k+1} -c.
\end{equation*}
ADMM converges under standard conditions, ensuring that both residuals—the primary residual $r = \norm{Ax^k + Bz^k - c}$ and the dual residual $s = \rho \norm{B^T(z^k-z^{k-1})}$—decrease to zero, signaling feasibility and stability. These residuals serve as indicators of the algorithm's convergence, and when both fall below predefined tolerances, ADMM is considered to have converged. This iterative process decomposes the problem into smaller subproblems, making ADMM well-suited for distributed computing environments \cite{boyd2011admm}.

OPF is a key problem in power system operations, aiming to determine optimal settings for control variables like generator outputs, voltage magnitudes, and reactive power injections to minimize an objective function while meeting physical and operational constraints. Common objectives include minimizing generation costs, power losses, or emissions. OPF is crucial for ensuring efficient and reliable power system operation, particularly as the grid evolves to accommodate renewable energy sources, dynamic loads, and decentralized energy resources. The problem is inherently non-convex due to the nonlinear power flow equations, making it computationally challenging to solve, particularly for large systems. Various formulations and solution techniques have been developed, including traditional AC-OPF and simplified DC-OPF models. Additionally, convex relaxations like semidefinite programming (SDP) and second-order cone programming (SOCP) offer computationally tractable approximations that provide near-optimal solutions. A detailed survey of these techniques is available in \cite{Molzahn2019}.

We apply ADMM to solve the OPF problem by decomposing it into subproblems. Leveraging the existing network structure, we divide the problem into its Transmission System Operator (TSO) and Distribution System Operator (DSO) components. The boundary buses between regions are duplicated, ensuring each TSO and DSO has its own copy of the relevant variables. Power flows through transformers are modeled differently in each region: in the TSO, power flows are treated as loads at boundary buses, while in the DSO, they are treated as generators. To ensure consistency across regions, we enforce the following equality constraints:
\begin{equation*}
p_i=-p_{i,\text{copy}},q_i=-q_{i,\text{copy}},V_i=V_{i,\text{copy}},\Theta_i=\Theta_{i,\text{copy}}
\end{equation*}
where $p_i$, $q_i$, $V_i$ and $\Theta_i$ represent active power, reactive power, voltage magnitude, and voltage angle at boundary bus $i$ in the TSO, and their counterparts $p_{i,\text{copy}}$, $q_{i,\text{copy}}$, $V_{i,\text{copy}}$ and $\Theta_{i,\text{copy}}$ represent the corresponding values in the DSO. Let $x$ denote the vector of variables in the TSO regions: 
\[
x^i := \begin{pmatrix}
V_i \\
\Theta_i \\
p_{i} \\
q_{i}
\end{pmatrix}, \text{for each boundary bus $i$ in the TSO} \]
and let $z$ represent the corresponding variables in the DSO:
 \[z^i :=\begin{pmatrix}
 V_{i,\text{copy}} \\
\Theta_{i,\text{copy}} \\
-p_{i,\text{copy}} \\
-q_{i,\text{copy}}
\end{pmatrix}, \text{ for each boundary bus $i$ in the DSO.}
\]
This setup allows independent optimization within each region while ensuring coordination at the boundaries. For ADMM applied to OPF, we use $A=I, B=-I, c=0$, simplifying the primary residual to $r^k=\|x^k - z^k\|$.  

\section{Vulnerabilities of the ADMM algorithm}
\label{sec: Vulnerabilities}
One method to detect tampering during ADMM iterations is to monitor the residual, which decreases as the algorithm converges, with significant deviations potentially indicating interference. However, if an attacker modifies the vector $z^k$ to $z_a^k = z^k + a$ by introducing an attack vector $a$, while ensuring that the residual 
remains unchanged, i.e. 
\begin{equation*}
\|x^k-z^k\|^2 = \|x^k - z_a^k\|^2,
\end{equation*}
the attack can bypass detection. The following theorem formalizes this condition. For simplicity, we may omit the superscript $k$ when the iteration is not relevant. 

\begin{thm}[Residual Evasion Criterion]
\label{Theorem undetectable definition}
If an attack vector $a$ satisfies the Residual Evasion Criterion
\begin{equation}
a^T(a - 2(x - z)) = 0
\end{equation} the residual $r$ retains its exact numerical value when $z$ is modified by the attack, i.e., when $z_a = z + a$.
\end{thm}

\begin{IEEEproof}
The claim follows directly from a straightforward computation: 
\begin{align*}
r_a^2 &= \|x - z_a\|^2 \\
&= \|x - (z + a)\|^2 \\
&= \|x - z\|^2 + \|a\|^2 - 2a^T(x - z) \\
&= \|x - z\|^2 + a^T(a - 2(x - z)) \\
&= \|x - z\|^2 = r^2.
\end{align*}
\end{IEEEproof}
With the condition for undetectable attacks established, the natural question is how to construct an attack vector $a$ that satisfies this condition. In the next section, we explore methods for identifying such vectors, addressing both feasibility and practical challenges.

\subsection{Constructing Random Undetectable Attack Vectors}
\label{sec: Constructing Random Undetectable Attack Vectors}
We demonstrate how to construct attack vectors $a$ that satisfy the Residual Evasion Criterion. Starting with a initial random vector, we use orthogonal decomposition to compute the attack vectors. Orthogonal decomposition expresses a vector as the sum of two components: one within a subspace and one orthogonal to it. Formally, for a vector $v$ and a subspace $W$, the decomposition is:
\begin{equation*}
v = v_\parallel + v_\perp,
\end{equation*}
where $v_\parallel \in \mathcal{W}$ and $v_\perp \perp \mathcal{W}$. This decomposition is uniquely defined and has widespread applications in optimization and numerical linear algebra \cite{strang2009introduction}. This method simplifies constructing the attack vector $a$ by splitting it into parallel and orthogonal components relative to $y = x - z$. We express $a = \lambda y + b$, where $\lambda$ is a scalar, $y$ is the parallel component, and $b$ is orthogonal to $y$. To find $b$, we use the Gram-Schmidt orthogonalization process (for a detailed introduction refer to \cite{strang2009introduction}), which ensures $b$ is orthogonal to $y$ by subtracting the projection of a vector $c$ onto $y$:
\begin{equation*}
c - \frac{c^T y}{y^T y} y.
\end{equation*}
After constructing a candidate for $b$, we scale it and choose $\lambda$ such that the Residual Evasion Criterion is satisfied:
\begin{align*}
a^T(a - 2y) &= (\lambda y + b)^T (\lambda y + b - 2y) = 0\\
 \iff \|b\|^2 &= -\lambda (\lambda - 2) \|y\|^2.
\end{align*}
With the relationship between $\lambda$, $b$, and $y$ firmly in place, we proceed with a specific example to illustrate their selection:

\begin{prop}
\label{exmple lambda = 1}
Setting
\begin{equation*}
\lambda = 1 
\end{equation*} 
and 
\begin{equation*}
b = \frac{\norm{y}}{\norm{c- \frac{c^Ty}{\norm{y}^2}y}} \left( c - \frac{c^Ty}{\norm{y}^2 }y \right)
\end{equation*}
yields a vector $a$ satisfying Theorem \ref{Theorem undetectable definition} for $y = x-z$.
\end{prop}
\begin{IEEEproof}
We have shown that the Residual Evasion Criterion is equivalent to:
\begin{equation*}
-(\lambda^2 - 2\lambda) \|y\|^2 = \|b\|^2.
\end{equation*}
For simplicity, we express $b$ as $b = \frac{\|y\|}{\|d\|} d$, where $d = c - \frac{c^T y}{\|y\|^2}y$.  Using this representation, we calculate:
\begin{equation*}
\|b\|^2 = \left\|\frac{\|y\|}{\|d\|} d\right\|^2 = \frac{\|y\|^2}{\|d\|^2} \|d\|^2 = \|y\|^2.
\end{equation*}
Substituting $\|b\|^2 = \|y\|^2$ back into the Residual Evasion Criterion and setting $\lambda = 1$ confirms the condition is satisfied. Since the vector $b$ is constructed using Gram-Schmidt, it is inherently orthogonal to $x-z$ by design. This can be easily verified with a straightforward calculation.  
\end{IEEEproof}

%Since the vector \(b\) is constructed using Gram-Schmidt, it is inherently orthogonal to \((x-z)\) by design. A simple calculation confirms this claim.
% However, to explicitly confirm this property and reinforce its role in ensuring the undetectability of the attack vector \(a = \lambda (x - z) + b\), we verify the orthogonality of \(b\) to \((x - z)\) as a consistency check.

%\begin{rem}
%The vector \(b\) is orthogonal to \((x-z)\).
%\end{rem}

%\begin{IEEEproof}
%We compute the scalar product \(b^T (x-z)\) as follows:
%\begin{align*}
%b^T (x-z) 
%&= \left( \frac{\|y\|}{\|c - \frac{c^Ty}{\|y\|^2} y\|} \left( c - \frac{c^Ty}{\|y\|%^2} y \right) \right)^T y \\
%&= \frac{\|y\|}{\|c - \frac{c^Ty}{\|y\|^2} y\|} \left( c^T y - \frac{c^Ty}{\|y\|^2} %\|y\|^2 \right) \\
%&= \frac{\|x-z\|}{\|c - \frac{c^T(x-z)}{\|x-z\|^2} (x-z)\|} \left( c^T (x-z) - c^T(x-z) \right) \\
%&= 0.
%\end{align*}
%Thus, \(b\) is orthogonal to \((x-z)\), as required.
%\end{IEEEproof}

\section{Goal-Oriented Attacks in ADMM}
\label{sec: Goal - Oriented Attacks in ADMM}
In this section, we move from adding random noise to a more targeted approach: designing attacks with specific goals, such as targeting particular variables or system components, while minimizing detection risk. Building on the previous example, we adapt our methodology to achieve these objectives and demonstrate the effectiveness of these strategies in comparison with existing ones on the IEEE 14-bus system in Section \ref{sec: Simulation Results}.

\subsection{Voltage Control as A Target}
\begin{figure}[!]
\centering
  \includegraphics[width=\linewidth]{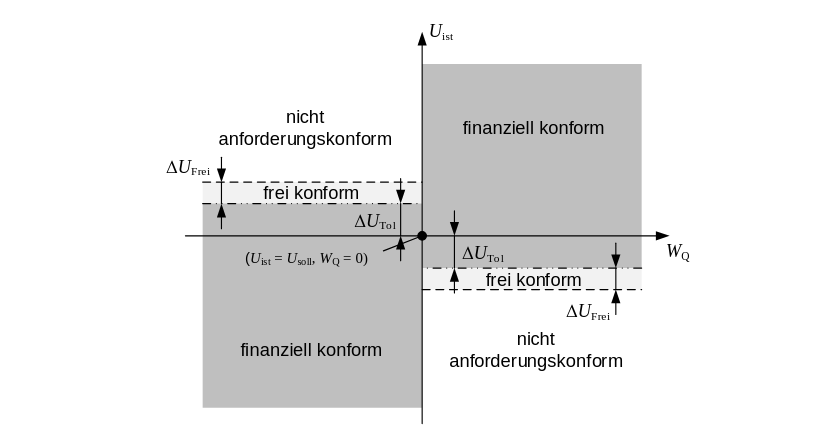}
  \caption{Voltage control for active role. $U_{\text{ist}}$ is the actual voltage and $U_{\text{soll}}$ is the target voltage at the feed-in node. $\Delta U_{\text{Tol}}$ is the billing tolerance, and $\Delta U_{\text{Frei}}$ is the free conformity band. $W_Q$ is the net reactive power exchange for the quarter-hour. The left side represents capacitance-like behavior (delivering reactive power), while the right side represents inductance-like behavior (consuming reactive power).  Source: Swissgrid \cite{swissgrid2020}} 
  \label{fig:voltage control}
\end{figure}

Voltage control is crucial for power system stability, making it a strategic target for adversarial attacks. Network participants manage reactive power to support voltage stability and receive financial incentives for compliance. Disrupting these mechanisms can have significant economic and operational consequences. Swissgrid, Switzerland's national TSO, oversees the high-voltage transmission grid and integrates it with the European network. It developed a voltage control framework to regulate reactive power exchange among power plants, distribution networks, and end-users, aiming to reduce system losses and improve efficiency. Updated in 2020, the system incentivizes compliance through financial rewards and penalties \cite{swissgrid2020}. Similar voltage control mechanisms are adopted by TSOs globally, including in Europe (e.g., TenneT, RTE, National Grid ESO), the United States (e.g., PJM, CAISO), and Asia (e.g., State Grid Corporation of China, Power Grid Corporation of India). These efforts highlight the universal necessity of voltage control for preventing equipment damage, optimizing power delivery, and maintaining stability. Swissgrid's framework incentivizes behavior within defined tolerance bands. Participants exchanging reactive power within these ranges are financially rewarded, while those outside are penalized. Active participants, like power plants, receive higher rewards, while semi-active participants, such as distribution networks and end-users, receive lower rewards. Compliance is ensured through real-time monitoring and monthly thresholds. 
For active participants, compensation and penalties are calculated based on the volume of reactive power exchanged and the deviation from the target voltage range, adjusted by predefined tolerance limits. Semi-active participants have a similar structure but include a free exchange zone around zero, where no charges apply. The tariff model is recalibrated annually based on historical data and projected costs, ensuring transparency and alignment with operational requirements \cite{swissgrid2019}. Fig. \ref{fig:voltage control} illustrates compliance zones:

\begin{itemize}
\item  Financially Compliant Zone (finanziell konform): Within this tolerance ($\Delta U_{\text{Tol}}$), reactive power exchanges are rewarded. The tolerance values are set at 1 kV for the 220-kV level and 2 kV for the 380-kV level.
\item  Free Compliant Zone: Beyond the compliant zone, within an additional tolerance ($\Delta U_{\text{Frei}}$, set at 1 kV for both voltage levels), exchange is neither compensated nor penalized. This zone allows technical flexibility but is not incentivized as it does not actively support the system.
\item Non-Compliant Zone: Exchanges outside the free zone incur penalties for failing to meet stability requirements.
\end{itemize}
    
In this section, we manipulate the set of variables $\lbrace z^i_4: i \in \mathfrak{A}\rbrace$, where $\mathfrak{A}$ represents the targeted boundary buses to influence financial rewards. The attacker uses a dual-pronged strategy targeting both voltage and reactive power at a boundary bus. The goal is to minimize payment by reducing $|z^i_4|$ while subtly pushing the network towards the voltage boundary for a brief period. This calculated maneuver leverages the natural fluctuations in the ADMM algorithm, which reduces the likelihood of detection. Simultaneously, the attack magnitude is kept small and the Residual Evasion Criterion is satisfied, both to avoid triggering alarms and to maintain the system's perceived stability. This can be formalized as: 
\begin{align}
\min \quad & \|a\|^2 + \sum_{i \in \mathfrak{A}} |z_4^i + a_4^i| \label{objective}\\ 
\text{s.t.} \quad & z_1^i + a_1^i - V_u \geq 0 \qquad \forall i \in \mathfrak{A} \label{voltage constraint}, \\
              & a^T (a - 2y) = 0 \label{undetectability constraint}.
\end{align}
where $z_1^i = V_i$, $z_4^i = -q_i$ and $V_u$ is the upper voltage limit. The objective \eqref{objective} minimizes the attack magnitude and payment, while \eqref{voltage constraint} pushes the system to the upper voltage boundary, and \eqref{undetectability constraint} is the Residual Evasion Criterion. While the voltage peak may seem suspicious, we show that the final voltage values remain well within the acceptable range. Furthermore, such transient peaks during the iterative computation process are entirely normal and can occur even in the absence of attacks. While our focus has been on the primal residual, the secondary residual can also indicate an attack. However, significant deviations in the secondary residual are rare when the primary residual remains unaffected. Nonetheless, if the attacker wishes to exercise additional caution, it is straightforward to incorporate safety measures into the optimization problem, for instance adding a constraint to ensure that the secondary residual does not exceed the average of the previous three iterations. Notably, the problem defined by \eqref{objective}–\eqref{undetectability constraint} can be efficiently addressed using standard off-the-shelf solvers.

\subsection{Random Attacks with Maximum Deviation}
Building on the previous section, we now examine the rationale behind setting $\lambda$ and $b$ as defined in Proposition \ref{exmple lambda = 1}, and explore how to construct attack vectors that maximize their disruptive impact. Rather than injecting arbitrary perturbations, a carefully designed attack can exploit the system's vulnerabilities while remaining undetected. By leveraging the orthogonal component $b$, we ensure that the attack influences the system in directions that significantly deviate from its expected trajectory, amplifying its disruptive effect. The equation $\norm{b}^2 = -(\lambda -1)^2 \norm{y}^2 + \norm{y}^2$ is a direct consequence of the Residual Evasion Criterion and highlights the trade-off between $\lambda$ and $b$, where maximizing one requires minimizing the other. Setting $\lambda=1$ maximizes the norm of the orthogonal component $b$, which can be computed using Gram-Schmidt and a random vector $c$: 
\begin{equation*}
b = \frac{\norm{y}}{\norm{c- \frac{c^Ty}{\norm{y}^2}y}} \left( c - \frac{c^Ty}{\norm{y}^2 }y \right).
\end{equation*}
Rather than using a random $c$, a more sophisticated attack explicitly defines $a$ and finds a suitable $c$ to construct $b$. By choosing $c$ already orthogonal to $x-z$, we simplify:
\begin{align*}
a = \lambda(x-z) + b =  (x-z) + \frac{\norm{x-z}}{\norm{c}} c.
\end{align*}
Rewriting this component-wise and squaring both sides yields:
%\begin{equation*}
%\frac{c_j}{\norm{c}} = \frac{a_j -( x_j - z_j)}{\norm{x-z}}
%\end{equation*}
%and squaring both sides yields
\begin{equation*}
\frac{c_j^2}{\norm{c}^2} = \frac{(a_j - x_j + z_j)^2}{\norm{x-z}^2}. 
\end{equation*}
If $a_j=x_j-z_j$, then $c_j=0$, but $c \neq 0$ must hold by definition of $a$. For simplicity, we assume $a_j \neq x_j-z_j$ for $j=1,\dots,4n$ where $n$ is the number of boundary buses. If this condition does not hold, the same reasoning applies to a subsystem containing only the subvector of $c$ where $c_j \neq 0$. Rearranging the terms results in: 
\begin{align*}
\left(1- \frac{\norm{x-z}^2}{(a_j-x_j+z_j)^2}\right)c_j^2 +  \sum_{\substack{i=1 \\ i \neq j}}^{4n} c_i^2 &= 0. 
\end{align*}
With a slight abuse of notation, define $A \in \mathbb{R}^{4n \times 4n}$ such that $A_{ij} = d_i$ if $i=j$ and $1$ if $i\neq j$, $y:= x-z$, $ d_i := 1 - \frac{\norm{y}^2}{(a_i-y_i)^2}$ and $c^2 := (
c_1^2, \cdots, c_{4n}^2 )^T$, we arrive at the following theorem:
\begin{thm}
\label{System of c}
If there exists a vector $c$ such that:
\begin{align*}
c^T y  = 0, \qquad 
A c^2  = 0, \qquad 
c \neq 0
\end{align*}
then it is possible to define an attack $a : = y + \frac{\norm{y}}{\norm{c}}c$ that satisfies the Residual Evasion Criterion (Theorem \ref{Theorem undetectable definition}). 
\end{thm}
Next, we determine when this system is solvable.
\begin{thm}
\label{Signs of c}
It is possible to define an attack  $a : = y + \frac{\norm{y}}{\norm{c}}c$ satisfying the Residual Evasion Criterion if for $j = 1,\dots,4n$ the signs of $c_j := \pm \frac{a_j-y_j}{\norm{y}}$ can be chosen such that $c^T y = 0$. 
\end{thm}
Before proving this, we introduce two helpful lemmas. Let $e$ denote the all-one vector of appropriate size. 
\begin{lem}
\label{Systemequalities}
For a matrix $A \in \mathbb{R}^{n \times m}$ there exists an $x$ satisfying 
\begin{equation}
\label{system 1}
\begin{aligned}
Ax =0, \qquad 
x  \geq 0, \qquad 
x  \neq 0
\end{aligned}
\end{equation}
if and only if there exists a $y$ such that  
\begin{equation}
\label{system 2}
\begin{aligned}
Ay =0, \qquad 
y \geq 0, \qquad 
e^Ty = 1.
\end{aligned}
\end{equation}
\end{lem}
\begin{IEEEproof}
Assume $x^*$ solves \eqref{system 1} and set $y^\star := \frac{x^\star}{e^T x^\star}$. Then 
\begin{align*}
Ay^\star &= A \frac{x^\star}{e^T x^\star} = \frac{1}{e^T x^\star} Ax^\star = \frac{1}{e^T x^\star} 0 = 0.
\end{align*} 
Since $x^\star \geq 0$ it follows that $y^\star \geq 0$. Additionally,  $e^Ty^* = 1$. Thus $y^*$ satisfies \eqref{system 2}. 
Conversely, if $y^*$ solves \eqref{system 2}, then $y^* \neq 0$ because $e^Ty^* = 1$ and $y^\star$ is also a solution to \eqref{system 1}.  
\end{IEEEproof}
We now proceed to the second lemma. 
\begin{lem}
\label{Lemma c orthogonal }
Let $A \in \mathbb{R}^{n \times n}$ be a matrix defined by 
\begin{align*}
A_{ij} = \begin{cases}
d_i, \text{ if } i = j, \\
1, \text{ if } i \neq j. 
\end{cases}
\end{align*}
There exists a vector $x$ satisfying \eqref{system 1}
if and only if $d_j <1$ for $j=1,\dots, n$ and 
\begin{align*}
\sum_{i=1}^{n} \frac{1}{1-d_i} = 1.
\end{align*}
Additionally, the solution is $x_j = \frac{1}{1-d_j}$ for $j=1,\dots, n$. 
\end{lem}
\begin{IEEEproof}
By Lemma \ref{Systemequalities} such an $x$ exists if and only if there is a $y$ satisfying \eqref{system 2}.  We show that these conditions hold for $y^\star_j := \frac{1}{1-d_j}$ for $j=1,\dots, n$. It is straightforward to verify that $y^*$ is also a solution to \eqref{system 1}. Since $d_j < 1$, we have $y^\star \geq 0$. 
Moreover, $e^T y^\star = \sum_{i=1}^n \frac{1}{1-d_i} =1 $. Now for $Ay^*$ 
\begin{align*}
(Ay^\star)_j = d_j y^\star_j + \sum_{\substack{i=1 \\ i \neq j}}^{n} y_i^\star
= d_j y_j^\star + \sum_{i=1}^n y_i^\star - y_j^\star 
= 0. 
\end{align*}
Conversely, assume feasibility and let $y^\star$ satisfy
\eqref{system 2}. 
From $Ay^\star =0$ and $e^T y^\star = 1$ we immediately get $y_j = \frac{1}{1-d_j}$. Since $y\geq 0$ this implies $d_j < 1$. Finally, $1 = e^Ty^\star$ gives $\sum_{i=1}^n \frac{1}{1-d_i}=1$ completing the proof. 
\end{IEEEproof}
We are now prepared for the proof of Theorem \ref{Signs of c}. 
\begin{IEEEproof}
First, we note that the subsystem 
\begin{align*}
Ac^2 &=0 \\
c & \neq 0 
\end{align*} is, due to the square, equivalent to system \eqref{system 1}. According to Lemma \ref{Lemma c orthogonal } this system is feasible if and only if $d_j <1$ for $j=1,\dots, n$ and $\sum_{i=1}^n \frac{1}{1-d_i}=1$. Since $d_j = 1- \frac{\norm{y}^2}{(a_j-y_j)^2}$ and $a = y + \frac{\norm{y}}{\norm{c}}c$, we immediately conclude that $d_j< 1$ because $y \neq 0$. Rearranging the terms in the definition of the attack vector $a$ gives the second condition:  
\begin{align*}
\norm{a-y}^2 = \norm{ \frac{\norm{y}}{\norm{c}}c}^2 = \frac{\norm{y}^2}{\norm{c}^2}\norm{c}^2 = \norm{y}^2
\end{align*}
and hence, 
\begin{align*}
\sum_{i=1}^n \frac{1}{1-d_i} = \sum_{i=1}^n \frac{(a_i-y_i)^2}{\norm{y}^2} = \frac{1}{\norm{y}^2}\norm{a-y}^2 = \frac{\norm{y}^2}{\norm{y}^2}=1.
\end{align*}
Therefore, following Lemma \ref{Lemma c orthogonal }, we obtain the unique solution 
\begin{align*}
c_j^2 = \frac{1}{1-d_j} = \frac{(a_j-y_j)^2}{\norm{y}^2}
\end{align*}
 and consequently, 
\begin{align*}
c_j = \pm \frac{a_j-y_j}{\norm{y}}.
\end{align*}
This means that if we can adjust the signs of $c_j$ such that $c^Ty=0$, we have identified a vector $c$ that satisfies the system in Theorem \ref{System of c}. As a result, this constructs an attack vector $a$ that avoids detection by monitoring ADMM residuals.
\end{IEEEproof}

\section{Simulation Results}
\label{sec: Simulation Results}
We conclude by demonstrating the effectiveness of our proposed attack strategies on the IEEE 14-bus system. First, we describe the experimental setup, followed by a detailed analysis of the results, examining how various parameters influence both the stealth and impact of the attacks.
\subsection{Experimental Setup}
\begin{figure}[!]
\centering
 \includegraphics[width = 0.5\textwidth]{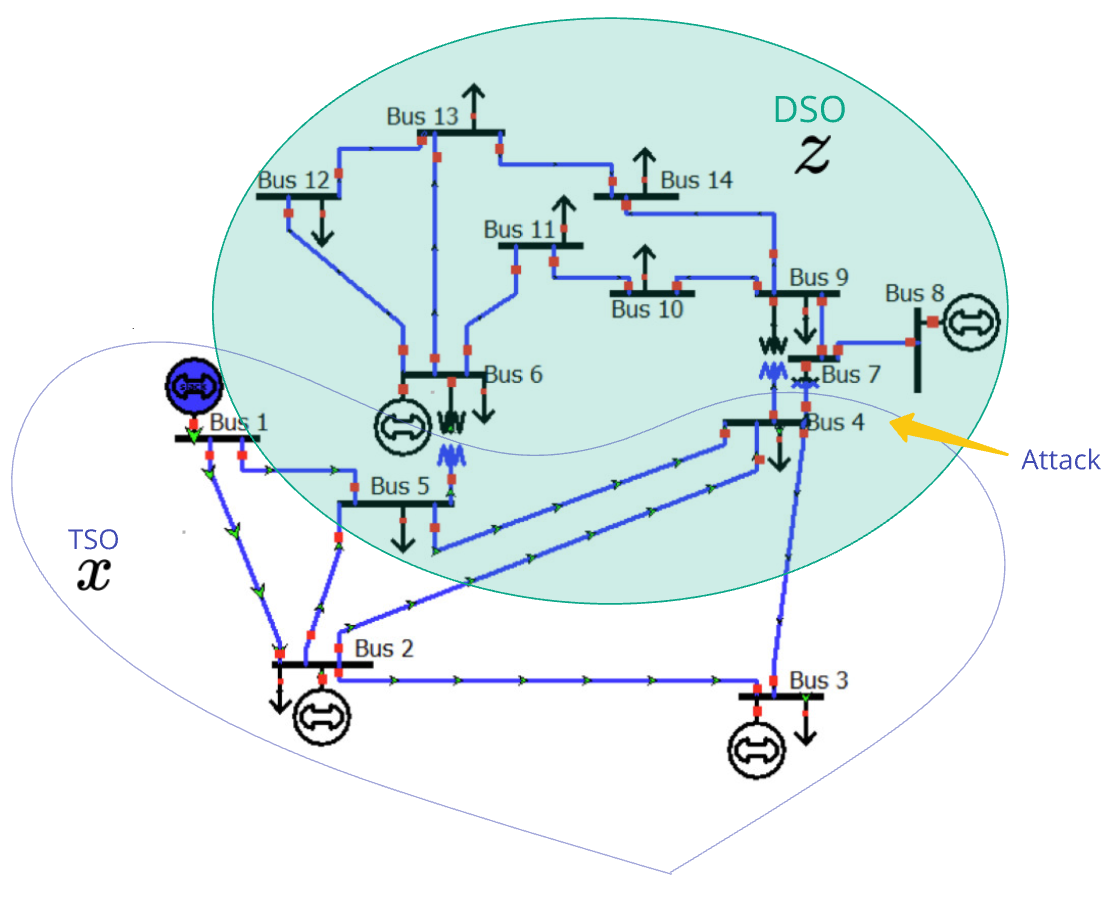}
  \caption{IEEE 14-bus test case and the split in TSO and DSO.}
  \label{fig: IEEE 14-bus case}
\end{figure}

The IEEE 14-bus test case, a simplified representation of the American Electric Power system as of February 1962, consists of 14 buses, 5 generators, and 11 loads, offering a practical model for analyzing power systems. 
Further details about the test case, including system configuration and parameters, can be found in  \cite{ieee14bus}. We partition the IEEE 14-bus network into a TSO part and a DSO part with the boundary defined at Bus 4 and Bus 5. Hence, duplicate Bus 4 and Bus 5, such that TSO and DSO both have their own copies to facilitate independent modeling. 
Then remove the transformers that originally connected the regions (i.e., transformers connecting Buses 5-6, 4-9, and 4-7). In the TSO network the power flows through these transformers are added to the loads at Buses 4 and 5, respectively.
In the DSO network, generators are introduced at Buses 4 and 5 to represent the power flows coming from the TSO. A visual representation is provided in Fig. \ref{fig: IEEE 14-bus case}. Lastly, we assume an upper voltage limit of $1.1$ p.u. and a lower voltage limit of $0.9$ p.u. 

We systematically evaluate several scenarios to investigate the impact and secrecy of different attack strategies. We start with naive attacks where a fixed percentage is added during a single iteration. We compare these simplistic attacks against the strategy described in Section \ref{sec: Constructing Random Undetectable Attack Vectors} explicitly focusing on Random Attacks with Maximum Deviation, i.e. Proposition \ref{exmple lambda = 1}. Finally, we turn our attention to the optimized attack vectors derived from problem \eqref{objective} - \eqref{undetectability constraint}, assessing their ability to balance stealth and effectiveness.
The attacks are restricted to boundary buses by design, with Bus 4 consistenly chosen as the target for controlled comparison across strategies. Each experiment involves a single attack during the ADMM calculation process to isolate its impact on system stability and performance. It is important to note that more powerful attacks could be devised by repeated targeting, increasing instability and manipulation. However, for the purposes of this study, we focus on single-step attacks to better understand their individual characteristics and detectability.

In the clean scenario, i.e., without cyber-attacks, the ADMM algorithm converges after 214 iterations. To evaluate the impact of attacks, we examine injections at different iterations. A complete list of attack scenarios can be found in Table \ref{tab:attack_scenarios}. To enhance reproducibility, we have also documented the specific attack vectors used in Table \ref{tab:Table attack vectors} in the Appendix. The attack vector refers to the perturbation introduced in an attacked iteration $k$, modifying the variable $z^k$ as $z^k_a = z^k+a$. The random vectors are generated using Python numpy.random.rand(), which produces random numbers sampled from a uniform distribution over the interval $\lbrack 0,1)$.  

\begin{table}[!t]
\caption{Attack scenarios at select iterations on boundary bus $4$ of TSO ($x$) or DSO ($z$). \label{tab:attack_scenarios}}
\centering
\begin{tabular}{|c||c||c||c||c|}
\hline
\textbf{Scenario} & \textbf{Type} & \textbf{Iteration} & \textbf{Atk. variable} \\ \hline
1 & Clean & - & - \\ \hline
2 & $+10\%$ in $q$ & 3 & $x$ \\ \hline
3 & $+3.5\%$ in $q$ & 3 & $x$ \\ \hline
4 & $-10\%$ in $q$ & 3 & $x$ \\ \hline
5 & $-3\%$ in $q$ & 3 & $x$ \\ \hline
6 & $+5\%$ in $q$ & 210 & $x$ \\ \hline
7 & $+3.5\%$ in $q$ & 210 & $x$ \\ \hline
8 & $+50\%$ in $q$ & 3 & $z$ \\ \hline
9 & $+100\%$ in $q$ & 3 & $z$ \\ \hline
10 & $+150\%$ in $q$ & 3 & $z$ \\ \hline
11 & $+250\%$ in $q$ & 3 & $z$ \\ \hline
12 & $+3\%$ in $q$ & 3 & $z$ \\ \hline
13 & $-50\%$ in $q$ & 3 & $z$ \\ \hline
14 & $+10\%$ in $V$, $-25\%$ in $q$ & 3 & $z$ \\ \hline
15 & \begin{tabular}[t]{@{}l@{}}$+10\%$ in $V$, $+50\%$ in $\Theta$, \\ 
$+20\%$ in $p$, $-25\%$ in $q$\end{tabular} & 3 & $z$ \\ \hline
16 & $+5\%$ in $q$ & 210 & $z$\\\hline
17 & $+3\%$ in $q$ & 210 & $z$\\\hline
18 - 27 & Proposition \ref{exmple lambda = 1} & 3 & $z$\\\hline
28 - 37 & Proposition \ref{exmple lambda = 1} & 20 & $z$\\\hline
38 - 47 & Proposition \ref{exmple lambda = 1} & 50 & $z$\\\hline
48 - 57 & Proposition \ref{exmple lambda = 1} & 100 & $z$\\\hline
58 - 67 & Proposition \ref{exmple lambda = 1} & 150 & $z$\\\hline
68 - 77 & Proposition \ref{exmple lambda = 1} & 210 & $z$\\\hline
78 & Problem \eqref{objective} - \eqref{undetectability constraint} & 3 & $z$ \\ \hline
79 & Problem \eqref{objective} - \eqref{undetectability constraint} & 20 & $z$ \\ \hline
80 & Problem \eqref{objective} - \eqref{undetectability constraint} & 50 & $z$ \\ \hline
81 & Problem \eqref{objective} - \eqref{undetectability constraint} & 100 & $z$ \\ \hline
82 & Problem \eqref{objective} - \eqref{undetectability constraint} & 150 & $z$ \\ \hline
83 & Problem \eqref{objective} - \eqref{undetectability constraint} & 210 & $z$ \\ \hline
\hline
\end{tabular}
\end{table}

\subsection{Results}
\begin{figure}[!]
\centering
  \includegraphics[width=\linewidth]{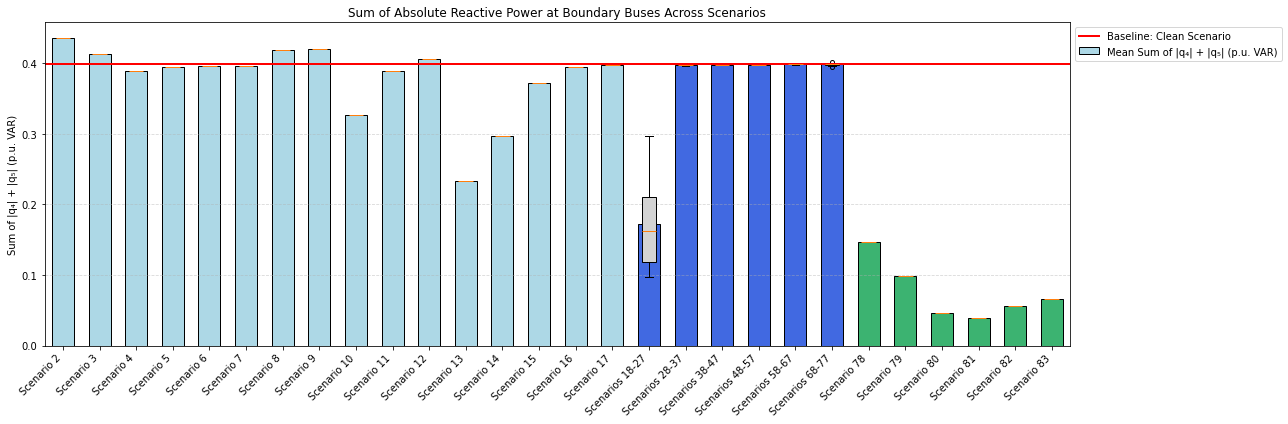}
  \caption{Financial impact of the attacks.}
  \label{fig:Figure Q abs all Scenarios}
\end{figure}

\begin{figure}[!]
\centering
  \includegraphics[width=\linewidth]{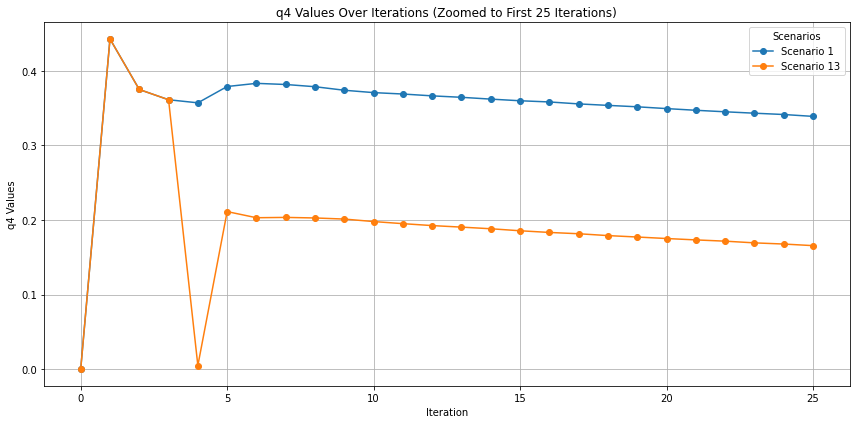}
  \caption{Reactive power at boundary bus 4 during ADMM, depicting a notable drop under attack.}
  \label{fig:Q values scenario 13}
\end{figure}

The financial reward depends on the total absolute reactive power at the boundary buses. Fig. \ref{fig:Figure Q abs all Scenarios} shows this value across scenarios, with a column plot for mean values and a boxplot for distribution. In most cases, the mean equals the attack value, except for attacks from Proposition \ref{exmple lambda = 1}, where the starting point is randomized. To account for variability, we averaged 10 trials. In the clean scenario, about $0.4$ VAR per unit is generated, covered by the TSO. Naive attacks struggle to balance impact and stealth; even a 50\% reduction in Scenario 13 has a smaller effect than Proposition \ref{exmple lambda = 1}. The sharp drop in reactive power at boundary bus 4 further underscores the limitations of such simple strategies (Fig. \ref{fig:Q values scenario 13}). Fig. \ref{fig:Figure Q abs all Scenarios} illustrates the influence of attack timing. The effectiveness of both the random undetectable attack (Proposition \ref{exmple lambda = 1}) and the optimization problem \eqref{objective}–\eqref{undetectability constraint} varies depending on the iteration at which the attack is applied. A clear trend emerges: later-stage attacks tend to have a weaker impact. This is due to Theorem \ref{Theorem undetectable definition}, which depends on the distance between $x$ and $z$, decreasing as the algorithm converges. This effect is especially pronounced in Proposition \ref{exmple lambda = 1}, where its influence diminishes beyond a certain iteration. In contrast, the optimization approach exhibits greater resilience, sustaining a more distributed impact across iterations, albeit still influenced by attack timing. Depending on the random starting point, Proposition \ref{exmple lambda = 1} can be more or less effective than the optimization problem \eqref{objective}–\eqref{undetectability constraint} or the naive attacks in the first 17 scenarios. Its key advantage is higher stealth while still achieving meaningful impact in many cases. Some random starting points may outperform the optimization approach, as the latter only approximates shifting the optimal power flow solution in a favorable direction. Scenario 22, a lower outlier in the boxplot, demonstrates that we have not identified the true optimal solution for the manipulated optimal power flow problem, as finding this solution would be excessively complex and computationally infeasible within a reasonable timeframe. 

\begin{figure}[!]
\centering
  \includegraphics[width=\linewidth]{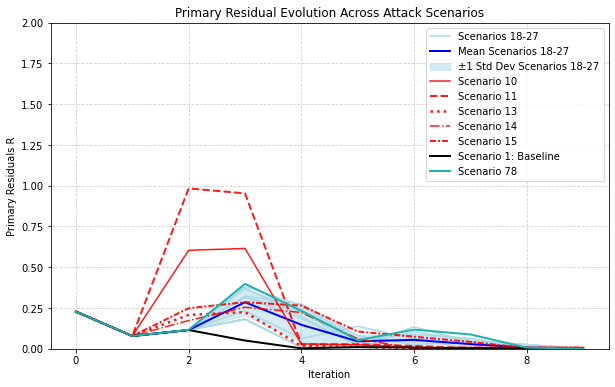}
  \caption{Zoomed-in view of the primary residual, highlighting noticeable peaks in naive attacks.}
  \label{fig:R residuals 10 iterations}
\end{figure}

\begin{figure}[!]
\centering
  \includegraphics[width=\linewidth]{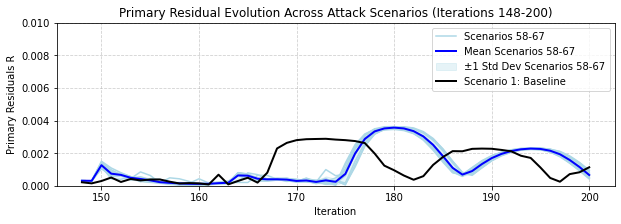}
  \caption{Zoomed-in view of the primary residual, demonstrating the stealth of the proposed attack.}
  \label{fig:R residuals 150 iterations}
\end{figure}

Fig. \ref{fig:R residuals 10 iterations} illustrates the primary residual for selected attack scenarios within the first 10 iterations. For randomized attacks, the mean residual (dark blue) is shown with individual scenarios (light blue) and a $\pm 1$ standard deviation shaded region. Colors indicate attack types as per the legend. This figure highlights a key flaw of the simple attacks: they fail to control the residual, causing detectable peaks. In contrast, our approach generates significantly smaller peaks that blend with normal fluctuations, making detection more difficult. Similarly, Fig. \ref{fig:R residuals 150 iterations} illustrates the trade-off between stealth and effectiveness. While random undetectable attacks (Scenarios 58–68) had little impact in Fig. \ref{fig:Figure Q abs all Scenarios}, their residual effects are nearly imperceptible. Note the plot's small scale necessary to even visualize these changes, which closely resemble the clean scenario (Scenario 1). Such attacks are valuable when stealth is paramount, enabling prolonged evasion of detection or subtle manipulation of system dynamics without raising alarms.

\begin{figure}[!]
\centering
  \includegraphics[width=\linewidth]{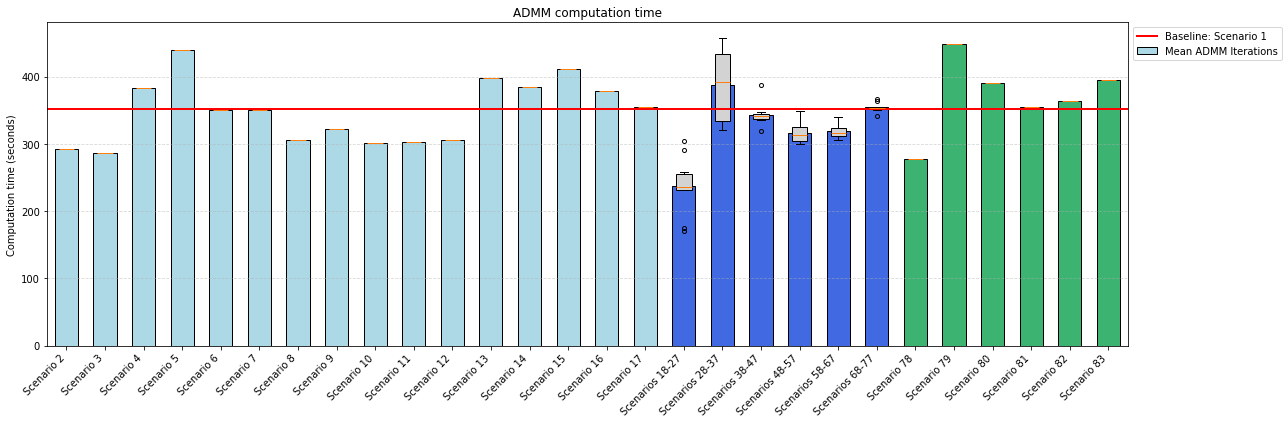}
  \caption{Stable ADMM computation time provides cover.}
  \label{fig: ADMM time}
\end{figure}

\begin{figure}[!]
\centering
  \includegraphics[width=\linewidth]{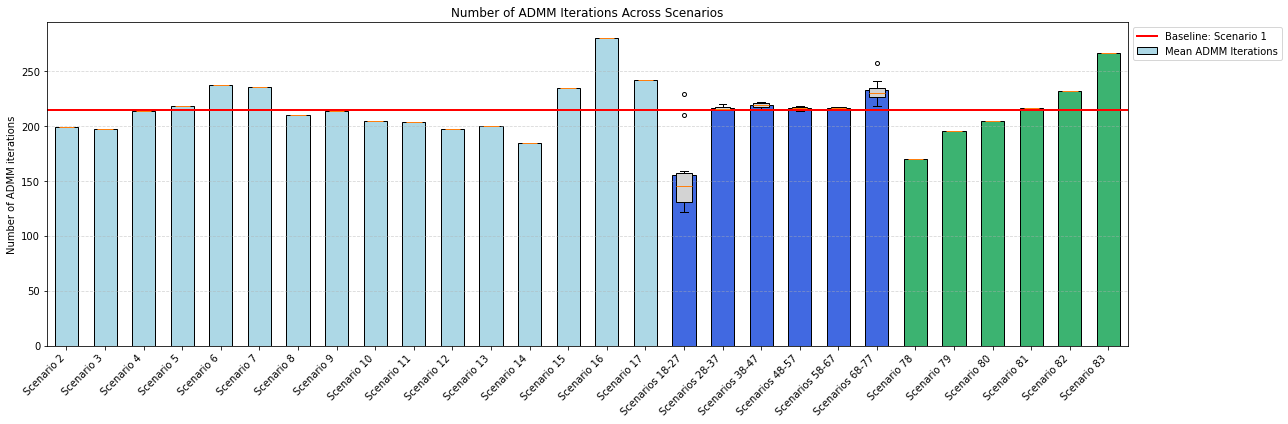}
  \caption{Consistency in ADMM iterations obscure detection.}
  \label{fig: ADMM iterations}
\end{figure}

Beyond residual monitoring, one might consider other ADMM parameters for attack detection, such as computation time or iteration count. However, as Fig. \ref{fig: ADMM time} and Fig. \ref{fig: ADMM iterations} show, these metrics remain remarkably stable across all attacks, providing a convenient cover. Attacks can effectively hide behind this stability, as their impact on these parameters is either negligible or, in some cases, even reduces iterations and computation time—though not enough to raise suspicion. While certain attacks are designed to explicitly disrupt these parameters, the ones studied here exploit their consistency to remain undetected.


\begin{figure}[!]
\centering
 \includegraphics[width=\linewidth]{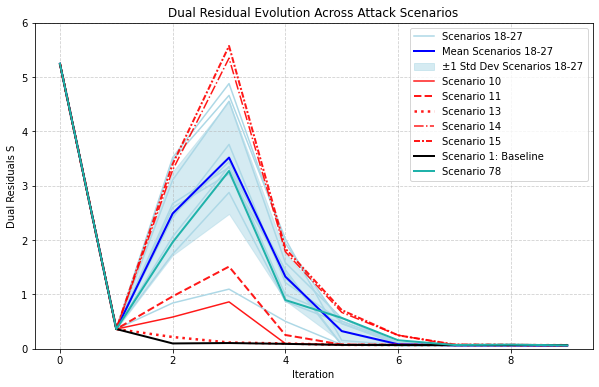}
  \caption{Zoomed-in dual residual, illustrating the advantage over naive strategies.}
  \label{fig: dual residual}
\end{figure}

While ADMM convergence is typically assessed using both the primal and dual residuals, our approach emphasizes the primal residual. This focus is justified, as the dual residual—although not directly controlled—in most cases remains within typical ranges and does not exhibit significant deviations that would undermine the reliability of the method, particularly in comparison to naive attack strategies. This observation is supported by Fig. \ref{fig: dual residual}, which also reveals that the dual residual often exceeds the primal residual. We focus on the first 10 iterations, where residual fluctuations are most pronounced before stabilizing. In a similar fashion as before, for Scenarios 18–27, the mean and $\pm 1$ standard deviation highlight the variability of randomized attacks. 

\begin{figure}[!]
\centering
  \includegraphics[width=\linewidth]{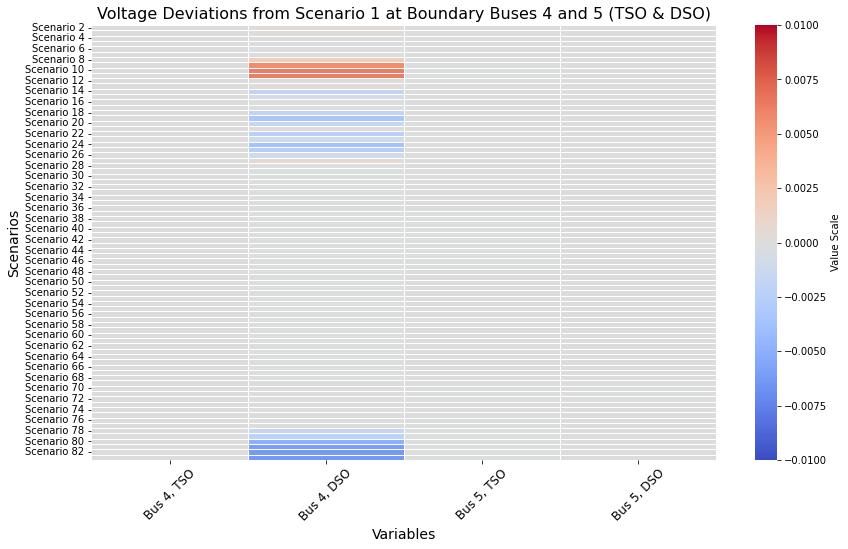}
  \caption{Heatmap of voltage deviations from the baseline, showing system stability and the potential to conceal attacks.}
  \label{fig: Voltage across scenarios}
\end{figure}

\begin{figure}[!]
\centering
  \includegraphics[width=\linewidth]{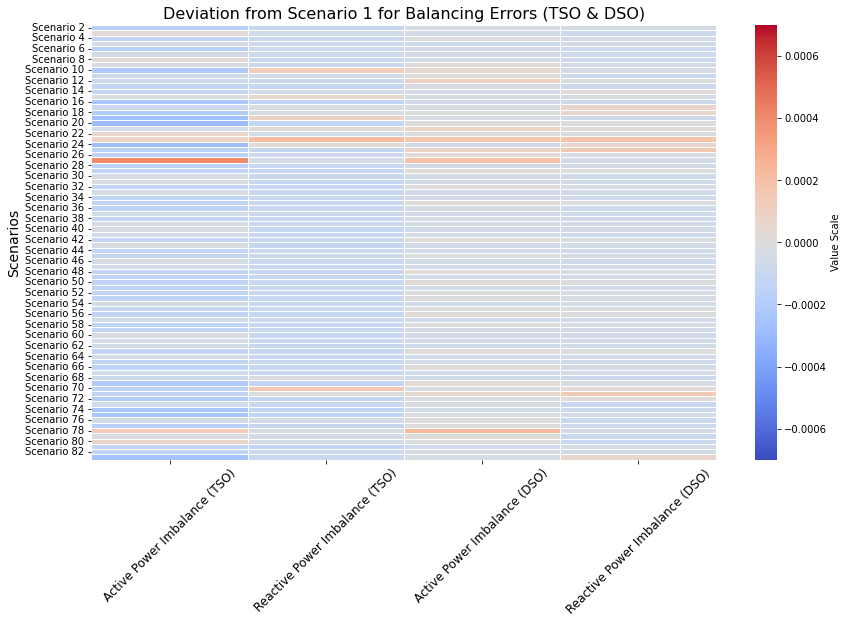}
  \caption{Heatmap of differences from baseline balancing errors, showcasing how attackers can exploit system stability.}
  \label{fig: balancing errors}
\end{figure}

Detecting attacks within the network itself, rather than through ADMM monitoring, is another potential approach. Voltage safety bounds are set at 1.1 and 0.9 p.u., yet Fig. \ref{fig: Voltage across scenarios} shows that even under random attacks, the system remains stable with no significant disturbances.  This robustness provides a strategic advantage for an attacker, as it allows manipulations to remain concealed within normal operating conditions. Interestingly, our optimization-based attack \eqref{objective}–\eqref{undetectability constraint} not only evades detection but even contributes to system stability, all while generating substantial financial gains for one party. Voltage deviations remain minor across all attack scenarios, ensuring the system stays within the financially favorable range, even with substantial $Q$ adjustments. Also on the network level, measuring balancing errors is a potential strategy for attack detection. In the context of optimal power flow, balancing errors are defined as the discrepancies between the computed power injections and the network's actual power demands. These errors emerge when the sum of generated power, minus the loads and network losses, fails to achieve the ideal condition of zero balance. However, our results in Fig. \ref{fig: balancing errors} show that the deviation of the balancing errors from the baseline remain minimal across all attack scenarios. This reinforces how the network's inherent stability can be leveraged by attackers to conceal their actions.

\section{Conclusion}
\label{sec: Conclusion}
In this paper, we designed two attack strategies that effectively evade detection by avoiding changes to the primary residual in the attacked iteration. Since many monitoring algorithms rely on residual analysis, this approach enables the attacks to bypass standard detection mechanisms. 

The first strategy uses a random starting point combined with Gram-Schmidt orthogonalization to ensure stealth. This approach can be refined by emphasizing the orthogonal component to maximize the disruption in the system. The second strategy extends the first one by allowing to target financial gains by simultaneously attacking reactive power and pushing the system to its upper voltage limit, exploiting the boundaries of operational constraints. An in-depth analysis and comparison of the two strategies highlighted the trade-offs between simplicity, detectability and effectiveness, providing valuable insights into the vulnerabilities of ADMM-based systems and the mechanisms by which they can be exploited. These findings emphasize the importance of designing more robust monitoring algorithms to protect against sophisticated attack strategies.

Future research could build on the foundation established in this paper by exploring attacks that target multiple buses or span multiple iterations. Coordinated strategies could leverage gradual, iterative adjustments to steer the system toward a desired state while remaining undetected. Additionally, while this paper employs a strict Residual Evasion Criterion to ensure stealth, future work could investigate the potential for relaxing this condition under practical assumptions, enabling the design of more adaptable and impactful attack strategies. These extensions would complement the findings of this study and contribute to a deeper understanding of system vulnerabilities. We conclude by noting that in Scenario 81, the total absolute reactive power remains nearly constant, yet boundary buses 4 and 5 show significant shifts, including sign changes. This suggests potential directions for future attack strategies to expand the scope of this study. 

In detection, many strategies focus on singular aspects, like monitoring residuals. However, a more robust algorithm could integrate multiple detection mechanisms. For example, in addition to residuals, analyzing changes in voltage, active power, and reactive power during ADMM iterations could help identify spikes indicative of an attack. This approach was exemplified in Fig. \ref{fig:Q values scenario 13}, where a naive attack left the residual unchanged but caused clear peaks in other parameters. Implementing multifaceted detection, however, presents challenges. Balancing false positives and false negatives is crucial, as setting the right threshold for detecting peaks is key to avoiding missed attacks without overloading the system. As detection strategies become more complex, careful calibration is needed to balance sensitivity and accuracy. Machine learning-based methods also offer promising potential for attack detection. Research, such as \cite{ReviewOfMLInPowerSystems} and \cite{Tuyizere2023}, shows how these techniques can detect patterns missed by traditional methods, providing an extra layer of security. Combining these methods with the insights from this paper could lead to a more resilient ADMM framework.

\section*{Acknowledgments}
The authors express their gratitude to Swissgrid for their support in this research. They extend special thanks to Ambra Toletti and Raphael Wu for their insightful discussions and valuable feedback, which greatly contributed to this work.

\appendix
\section{Appendix}
\noindent 
\FloatBarrier 
\centering
\footnotesize
\captionof{table}{Attack Vector Values}
\tablehead{
    \hline
    \textbf{Scenario} & \textbf{Attack Vector} \\
    \hline
}

\tabletail{
    \hline
    \multicolumn{2}{r}{\textit{Continued on next page...}} \\
}

\tablelasttail{
    \hline
    \multicolumn{2}{r}{\textit{End of table.}} \\
}
\begin{supertabular}{|c||l|}
\label{tab:Table attack vectors}
18 & [0.02121068 0.0130993  0.12314844 0.0864099 ]\\ \hline
19 &  [0.07341862 0.06317238 0.11244587 0.03505106]\\ \hline
20 & [0.05369313 0.00871949 0.12110739 0.07501809]\\ \hline
21 &  [0.01541827 0.07886334 0.11617012 0.05745474]\\ \hline
22 & [0.06255644 0.01532933 0.11979587 0.06895177]\\ \hline
23 & [0.05273261 0.09606387 0.10602196 0.00201061]\\ \hline
24 &  [0.06933667 0.02945763 0.11788314 0.06067893]\\ \hline
25 & [0.04830191 0.05887625 0.11715014 0.06107415]\\ \hline
26 &[0.01850385 0.08056118 0.11569358 0.05510166]\\ \hline
27 &  [0.03478083 0.10240982 0.10712628 0.00895416]\\ \hline 
28 & [6.52923e-05  1.11296e-04  9.81143e-05 -1.25645e-04]\\ \hline
29 &  [8.28166e-05  1.71912e-05  1.57168e-04 -1.01030e-04]\\ \hline
30 & [2.33227e-05  9.23689e-05  1.45938e-04 -1.08116e-04] \\ \hline
31 & [6.57201e-05  9.67342e-05  1.21174e-04 -1.17061e-04] \\ \hline 
32 & [1.17634e-04  7.47446e-05  6.61172e-05 -1.35156e-04] \\ \hline
33 & [8.55322e-05  8.13466e-05  1.20480e-04 -1.16682e-04] \\ \hline 
34 & [8.27203e-05  8.71954e-05  1.16819e-04 -1.18198e-04] \\ \hline 
35 & [3.01434e-06  7.89714e-05  1.58809e-04 -1.02943e-04] \\ \hline 
36 & [5.52432e-05  1.04791e-04  1.18457e-04 -1.18312e-04]\\ \hline 
37 & [1.28545e-04  4.18002e-05  8.94557e-05 -1.25657e-04]\\ \hline 
38 &  [0.00033554 0.00060707 0.00127095 0.00013236]\\ \hline 
39 & [6.06571e-04 4.56323e-04 1.23767e-03 7.72225e-05]\\ \hline 
40 &  [0.0001615   0.00090441  0.00112068 -0.00011829]\\ \hline 
41 & [0.0004926   0.00083299  0.00106443 -0.00021075]\\ \hline 
42 & [0.00025234 0.00065363 0.00126768 0.00012666]\\ \hline 
43 & [8.81615e-04  3.67449e-05  1.15372e-03 -6.53645e-05]\\ \hline 
44 & [5.56211e-04 5.30295e-04 1.23234e-03 6.85818e-05]\\ \hline 
45 & [0.00024766 0.00066986 0.00126115 0.00011582]\\ \hline 
46 & [6.42585e-04  6.34219e-04  1.13597e-03 -9.14259-05]\\ \hline 
47 & [0.00028043 0.00027531 0.00136882 0.00029301]\\ \hline 
48 & [1.10091e-04  7.24838e-05 -4.05319e-05 -1.30556e-04]\\ \hline 
49 & [7.70328e-05  1.04375e-04 -1.01420e-04 -9.45919e-05]\\ \hline 
50 & [3.68833e-05  1.29133e-04 -8.66583e-05 -1.02549e-04]\\ \hline 
51 & [4.36177e-05  1.26842e-04 -4.52692e-05 -1.26574e-04]\\ \hline 
52 & [8.86995e-05  1.02790e-04 -6.93971e-05 -1.13192e-04]\\ \hline 
53 & [7.31012e-05  9.23426e-05 -8.29755e-06 -1.48735e-04]\\ \hline 
54 & [1.51510e-05  8.37389e-05  2.48265e-05 -1.67940e-04]\\ \hline 
55 & [5.45215e-05  1.22520e-04 -8.69528e-05 -1.02553e-04]\\ \hline 
56 & [9.88206e-05  8.50837e-05 -3.48270e-05 -1.33582e-04]\\ \hline 
57 & [1.14445e-04  6.52528e-05 -4.08441e-05 -1.30529e-04]\\ \hline 
58 & [1.52442e-04  6.38828e-05  2.64114e-04 -1.48861e-04]\\ \hline 
59 & [0.00013494  0.00018196  0.00018477 -0.00018379]\\ \hline 
60 & [2.31975e-04  5.33502e-05  1.48807e-04 -2.01080e-04]\\ \hline 
61 & [0.00010482  0.00014033  0.00025586 -0.00015195]\\ \hline 
62 & [1.73608e-04  3.61746e-05  2.53083e-04 -1.54066e-04]\\ \hline 
63 & [1.47308e-05  1.81783e-04  2.49261e-04 -1.54410e-04]\\ \hline 
64 & [1.91624e-04  9.71710e-05  2.06146e-04 -1.74863e-04]\\ \hline 
65 & [1.72501e-04  7.49430e-05  2.42232e-04 -1.58701e-04]\\ \hline 
66 & [0.00010181  0.00021101  0.00016562 -0.00019213]\\ \hline 
67 & [1.41356e-04  5.52496e-05  2.74632e-04 -1.44149557e-04]\\ \hline 
68 & [8.19125e-05 4.59841e-04 1.08182e-03 5.64933e-04]\\ \hline 
69 & [0.00060425 0.00040423 0.0010303  0.00034295]\\ \hline
70 & [0.00050432 0.00034494 0.00105841 0.0004627 ]\\ \hline 
71 & [0.00075692 0.0002196  0.00101032 0.00025642]\\ \hline 
72 & [5.461385e-05 3.43557e-04 1.09533e-03 6.21999e-04]\\ \hline 
73 & [0.00035637 0.00059902 0.00103831 0.00037914]\\ \hline 
74 & [4.72707e-04 9.57743e-05 1.08032e-03 5.54913e-04]\\ \hline 
75 & [5.74563e-04  7.05504e-04  9.36370e-04 -5.47820e-05]\\ \hline 
76 & [0.00025254 0.00075649 0.00100612 0.00024358]\\ \hline 
77 & [0.00073719  0.00054561  0.00092492 -0.00010501]\\ \hline 
78 & [0.06150997  0.0029498   0.08841819 -0.10452701]\\ \hline 
79 & [6.02433e-02 -6.51130e-06  4.28351e-05 -4.23917e-02]\\ \hline 
80 & [5.78314e-02  4.18136e-06  8.05939e-04 -4.40490e-02]\\ \hline 
81 & [5.66638e-02  2.08856e-06 -6.12412e-05 -4.42714e-02]\\ \hline 
82 & [5.64804e-02  1.22251e-06  9.13973e-05 -4.44684e-02]\\ \hline 
83 & [5.64435e-02 -2.77430e-09  9.01191e-04 -4.44423e-02] \\ \hline
\hline
\end{supertabular}

\section{Biography Section}

\vspace{11pt}

%\bf{If you include a photo:}\vspace{-33pt}
%\begin{IEEEbiography}[{\includegraphics[width=1in,height=1.25in,clip,keepaspectratio]{fig1}}]{Michael Shell}
%Use $\backslash${\tt{begin\{IEEEbiography\}}} and then for the 1st argument use $\backslash${\tt{includegraphics}} to declare and link the author photo.
%Use the author name as the 3rd argument followed by the biography text.
%\end{IEEEbiography}
%\begin{IEEEbiographynophoto}{Sabrina Bruckmeier}

\begin{IEEEbiographynophoto}{Sabrina Bruckmeier}
\justifying
is a Ph.D. student in discrete optimization at ETH Zurich. She holds a Bachelor's degree in Industrial Mathematics, which combines Electrical Engineering, Computer Science, and Mathematics, as well as a Master’s degree in Mathematics with honors from Friedrich Alexander University Erlangen-Nürnberg. Her research focuses on robust optimization in power systems, sparse approximation in signal processing and machine learning, and flows over time in hypergraphs. Sabrina’s interests include mathematical optimization, algorithm design, and cybersecurity in critical infrastructure.
\end{IEEEbiographynophoto}

\begin{IEEEbiographynophoto}{Dr. Huadong Mo}
\justifying
is a Senior Lecturer at the University of New South Wales in Australia and Coordinator of the Systems Enineering Discipline within the School of Systems and Computing. Previously, he was a Postdoctoral Fellow at the Swiss Federal Institute of Technology, Zurich. He earned his Ph.D. from the City University of Hong Kong. His research focuses on enhancing the reliability, resilience, and security of complex systems using learning-based algorithms, with applications in power systems, cyber-physical systems, and manufacturing. He has published over 80 SCI-indexed papers and secured approximately 5 million AUD in research funding. A recipient of the 2024 IEEE SMC Early Career Award—the fourth since its inception—Dr. Mo was also awarded the 2023 Visiting Research Fellowship (Pre-award of the Jean d'Alembert Pour Fellowship) in France. He is a Senior Member of IEEE, Chair of the IEEE SMC ACT Chapter, and serves on the editorial boards of multiple SCI-indexed journals.
\end{IEEEbiographynophoto}

\begin{IEEEbiographynophoto}{James Ciyu Qin}
\justifying
received a BE (Hons) degree in Mechanical and Manufacturing Engineering and a PhD in Systems Engineering from the University of New South Wales, Australia, in 2019 and 2024, respectively.
He is currently a Postdoctoral Researcher at the Reliability and Risk Engineering Laboratory, Institute of Energy Technology, ETH Zürich, Switzerland. His primary research focuses on enhancing the resilience, performance, and security of complex systems through various robust optimisation techniques. His research interests include developing preventive and corrective procedures that account for the risk of system failures and providing risk-aware prevention measures for resilience enhancement.
\end{IEEEbiographynophoto}

\vfill
\bibliographystyle{IEEEtran}
\bibliography{bibliography} 
\end{document}